\documentclass{iopart}
\usepackage{graphicx}

\expandafter\let\csname equation*\endcsname\relax
\expandafter\let\csname endequation*\endcsname\relax
\usepackage{amsmath,amssymb,amsbsy,amsfonts}
\usepackage{natbib}
\usepackage[british]{babel}
\usepackage{units}
\usepackage{hyperref}
\usepackage{pdflscape}

\newcommand{\eq}[1]{Eq.\,\eqref{#1}}
\newcommand{\eqs}[1]{Eqs.\,\eqref{#1}}
\newcommand{\cc}{\mathrm{c}}

\newcommand{\ud}{\mathrm{d}}
\newcommand{\G}{\mathrm{G}}
\newcommand{\rg}{r_{\text{\tiny$g$}}}
\newcommand{\rQ}{q}

\newcommand{\s}[1]{{\text{$#1$}}\hspace{-3pt}}

\begin{document}

\title[Newtonian description of particle motion in spherically symmetric spacetimes]{Generalized Newtonian description of particle motion in spherically symmetric spacetimes}

\author{Emilio Tejeda and Stephan Rosswog}
\address{Department of Astronomy and Oskar Klein Centre, Stockholm University, AlbaNova, SE-10691 Stockholm, Sweden}
\ead{emilio.tejeda@astro.su.se}

\begin{abstract}
We present a generalized Newtonian description of particle dynamics valid for any spherically symmetric, static black hole spacetime. This approach is derived from the geodesic motion of test particles in the low-energy limit. It reproduces exactly the location of the marginally stable, marginally bound, and photon circular orbits; the radial dependence of the energy and angular momentum of circular orbits; parabolic motion; pericentre shift; and the spatial projection of general trajectories. As explicit examples of the new prescription, we apply it to the Schwarzschild, Schwarzschild--de Sitter, Reissner--Nordstr{\"o}m, Ay{\'o}n-Beato--Garc{\'i}a, and Kehagias--Sfetsos spacetimes. In all of these examples, the orbital and epicyclic frequencies are reproduced to better than $10\%$.  The resulting equations of motion can be implemented easily and efficiently within existing Newtonian frameworks. 
\end{abstract}

\section{Introduction}

Since its birth almost 100 years ago, the theory of general relativity (GR) has successfully withstood continuous experimental testing: from the anomalous perihelion precession of Mercury and the bending of light rays from distant stars by the Sun to the indirect confirmation of the existence of gravitational waves from the orbital decay of the Hulse--Taylor pulsar \citep{will06}. All of these examples, however, only probe the weak field limit of the theory while the strong field regime remains largely uncharted \citep{psaltis}. One of the most dramatic consequences of the strong field limit of GR is the prediction of black holes and, although current astrophysical observations are consistent with their existence, the available evidence does not allow to discriminate between the objects predicted by GR and those coming from alternative gravitational theories \citep[see, e.g.][]{barausse13,bambi13}.

It is generally believed that astrophysical black holes possess a certain degree of intrinsic angular momentum, either since birth or as a result of subsequent accretion processes \citep{narayan}. For this reason, it is expected that the spacetime around astrophysical black holes should be well approximated by the solution of a rotating black hole found by \cite{kerr}. Nevertheless, many key relativistic features can already be investigated in the non-rotating, spherically symmetric black hole solution of Schwarzschild. Moreover, one of the first steps in the development of an alternative theory of gravity is to look for the kind of black hole solutions that it admits, and, given their simplicity, a prominent role in this search is played by static, spherically symmetric black hole spacetimes.

The Schwarzschild black hole solution of GR is the best known example of a static and spherically symmetric spacetime, but there exist other metrics that might be relevant for different astrophysical scenarios, especially in the context of alternative theories of gravity and/or extra degrees of freedom for the central object. Examples include
\begin{itemize}
\item[{\bf-}] The Reissner--Nordstr\"om solution describing a charged black hole within Einstein--Maxwell equations. For a discussion of particle motion in this spacetime see, e.g~\cite{bicak89,pugliese,grunau}.
\item[{\bf-}] The Schwarzschild--de Sitter spacetime describing a Schwarzschild black hole in an expanding universe with cosmological constant. See \cite{stuchlik08,hackmann08b,hackmann08a} for a general discussion of the motion of test particles in this spacetime.
\item[{\bf-}] The Kehagias--Sfetsos spacetime \citep{kehagias}, a black hole solution of the Lorentz-violating gravitational theory of \cite{horava}. The motion of test particles in this spacetime has been discussed in, e.g.~\cite{abdu,enolskii,vieira}. 
\item[{\bf-}] The Boulware--Deser black hole solution in (4+1)-dimensional Gauss--Bonnet gravity \citep{boulware}.
\item[{\bf-}] Regular black holes. These are curvature singularity-free, exact solutions of Einstein equations coupled to some nonlinear electrodynamics \citep[see, e.g.][]{bardeen68,ABG}. See \cite{zhou,garcia} for a discussion of test particle motion in these spacetimes.
\item[{\bf-}] A Schwarzschild black hole pierced by a cosmic string \citep{aryal}. For a discussion of geodesic motion in this spacetime see, e.g.~\cite{hackmann}.
\item[{\bf-}] Schwarzschild and Reissner--Nordstr\"om black holes in spacetimes with higher dimensions \citep{tangherlini,emparan}. For a general discussion of geodesic motion in these spacetimes see, e.g.~\cite{hackmann08}.
\end{itemize}  

In this article we present a generalized Newtonian description of the motion of test particles in any given static, spherically symmetric spacetime. This work extends the scheme of \cite{TR} (referred to as Paper~I in the following) where, by considering the low-energy limit of geodesic motion, we introduced an  accurate Newtonian description of particle motion around a Schwarzschild black hole. The approach presented in Paper~I reproduces exactly certain key relativistic features of this spacetime (e.g.~radial location, angular momentum and energy of distinct circular orbits), while several other properties (e.g.~Keplerian and epicyclic frequencies) are described with a better accuracy than with commonly used pseudo-Newtonian potentials. Moreover, it showed a very good agreement with the thin disc accretion model of \cite{novikov73} and the analytic model for relativistic accretion of \cite{tejeda2,tejeda3}.

The paper is organized as follows. In Section~\ref{S2} we present the model and discuss the motion of general test particles within the new approach. In Section~\ref{S3} we derive explicit expression for the equations of motion that can be easily implemented within existing Newtonian frameworks. Circular orbits are discussed in detail in Section~\ref{S4} while perturbations away from them are considered in Section~\ref{S5}. Finally, we summarize our results in Section~\ref{S6}.

\section{Generalized Newtonian model}
\label{S2}

In general, we can express the differential line element in a static, spherically symmetric spacetime as
\begin{gather}
\ud s^2 = -\alpha\,\cc^2\ud t^2 + \alpha^{-1}\,\ud r^2+ r^2 \ud \theta^2 +r^2 \sin^2\theta\,\ud \phi^2 ,\label{e1}\\
\alpha = 1+2\,\Phi/\cc^2, \nonumber
\end{gather}
where c is the speed of light and $\Phi$ is a function of the radial coordinate $r$ only. 

The staticity and spherical symmetry of the metric in \eq{e1} imply that the orbit of a test particle is confined to a single plane (orbital plane) and that its motion is characterized by the existence of two conserved quantities, the specific energy $\mathcal{E}$ and the specific angular momentum $h$. They are given by
\begin{gather}
\mathcal{E} = \alpha\,\cc^2\frac{\ud t}{\ud \tau}, 
\label{e2}\\
h = r^2 \frac{\ud \varphi}{\ud \tau},
\label{e3}
\end{gather}
where $\tau$ is the proper time and $\varphi$ is an angle measured within the orbital plane. The proper time is related to the proper distance by \mbox{$\ud s^2 = -\cc^2 \ud \tau^2$}, while the angle $\varphi$ is related to $\theta$ and $\phi$ through $\ud \varphi^2 = \ud \theta^2 +\sin^2\theta\,\ud \phi^2 $. 

In addition to the specific relativistic energy defined in \eq{e2}, we introduce the specific mechanical energy defined as
\begin{equation}
E = \frac{\mathcal{E}^2-\cc^4}{2\,\cc^2}.
\label{e4}
\end{equation}
By combining \eqs{e1} and \eqref{e2} we get
\begin{equation}
\frac{\mathcal{E}^2}{\cc^4} \left( \cc^2 - \frac{\dot{r}^2}{\alpha^2}-\frac{r^2\dot{\varphi}^2}{\alpha}\right) = \alpha\,\cc^2 ,
\label{e5}
\end{equation}
where a dot denotes differentiation with respect to the coordinate time $t$. Using \eq{e4} and considering the low-energy limit of \eq{e5}, in which $\mathcal{E}\simeq\cc^2$ or, equivalently, $|E| \ll \cc^2$, we obtain 
\begin{equation}
E \xrightarrow[\mathcal{E}\simeq\cc^2]{} E_* \equiv 
\frac{1}{2}\left(\frac{\dot{r}^2}{\alpha^2}+\frac{r^2\dot{\varphi}^2}{\alpha}\right) +\Phi.
\label{e6}
\end{equation}
The symbol $*$ is used here and in the following to indicate that the corresponding quantity is being evaluated in the low-energy limit and, thus, to distinguish it from the exact relativistic value.

The low-energy limit of the specific mechanical energy $E_*$, as defined in \eq{e6}, is the basis of the present generalized Newtonian description, since, if we introduce the Lagrangian 
\begin{equation}
L = \frac{1}{2}\left(\frac{\dot{r}^2}{\alpha^2}+\frac{r^2\dot{\varphi}^2}{\alpha}\right) -\Phi,
\label{e7}
\end{equation}
it is simple to verify that $L$ is connected to $E_*$ through the usual expression 
\begin{equation}
E_* = \dot{r}\frac{\partial L}{\partial \dot{r}}+
\dot{\varphi}\frac{\partial L}{\partial \dot{\varphi}}-L
 = \frac{1}{2}\left(\frac{\dot{r}^2}{\alpha^2}+\frac{r^2\dot{\varphi}^2}{\alpha}\right) +\Phi.
\label{e8}
\end{equation}
On the other hand, the independence of $L$ from the orbital angle $\varphi$ leads to the conservation of the specific angular momentum 
\begin{equation}
h_* = \frac{\partial L}{\partial \dot{\varphi}} = 
\frac{r^2\dot{\varphi}}{\alpha}.
\label{e9}
\end{equation}
In full agreement with the low-energy limit, $h_*$ is related to the relativistic angular momentum as $h_* = (\cc^2/\mathcal{E})h$.

Note that the Lagrangian in \eq{e7} can be rewritten as
\begin{gather}
L = T - \Phi_\G,\\ \Phi_\G = \Phi\left[1 + \frac{(1+\alpha)\dot{r}^2}{\alpha^2\cc^2}
+\frac{r^2\dot{\varphi}^2}{\alpha\,\cc^2}\right] ,
\end{gather}
where $T = (\dot{r}^2+r^2\dot{\varphi}^2)/2$ is the specific kinetic energy in non-relativistic physics and $\Phi_\G$ is the extension of the generalized (velocity-dependent) potential introduced in Paper I.

By combining \eqs{e8} and \eqref{e9} we obtain the equation governing the radial motion 
\begin{equation}
\dot{r} = \pm\alpha\sqrt{2\,E_*-2\,\Phi-\alpha\,\frac{h_*^2}{r^2}},
\label{e10}
\end{equation}
which, up to the constant factor $\mathcal{E}/\cc^2\simeq1$, coincides with the corresponding relativistic equation that one obtains from \eqs{e3}, \eqref{e4} and \eqref{e5}. Note in particular that \eq{e10} coincides exactly with the relativistic expression for parabolic-like trajectories for which $\mathcal{E} = \cc^2$.

From \eqs{e9} and \eqref{e10} we see that the spatial projection of a general test particle trajectory is described in terms of the integral
\begin{equation}
\varphi = \int \frac{h_*\,\ud r}{r\sqrt{2(E_*-\Phi)r^2-\alpha\,h_*^2}},
\label{e11}
\end{equation}
which, after making the correspondences $E_*\leftrightarrow E$ and $h_*\leftrightarrow h$, is formally identical to the exact relativistic expression. This implies that both the geometrical path and the pericentre shift of general trajectories are captured exactly by the present generalized Newtonian description.

\section{Equations of motion}
\label{S3}

The equations of motion in terms of a general reference frame $(r,\ \theta,\ \phi)$ are calculated by applying the Euler--Lagrange equations to the Lagrangian in \eq{e7}. The resulting expressions are
\begin{align}
\ddot{r}  = & -\alpha^2 \Phi'
+ \frac{\alpha'}{\alpha}\,\dot{r}^2 + \alpha\,r\left(1-\frac{r\,\alpha'}{2\,\alpha}\right)
\left(\dot\theta^2 + \sin^2\theta\,\dot\phi^2\right) , 
\label{e12} \\
\ddot{\theta} = &
-\frac{2\,\dot{r}\,\dot{\theta}}{r}\left(1-\frac{r\,\alpha'}{2\,\alpha}\right)
+\sin\theta\cos\theta\,\dot\phi^2,
\label{e13}  \\
\ddot{\phi} = &
-\frac{2\,\dot{r}\,\dot{\phi}}{r}\left(1-\frac{r\,\alpha'}{2\,\alpha}\right)
-2\,\cot\theta\,\dot\phi\,\dot\theta,
\label{e14} 
\end{align}
where a prime denotes the derivative with respect to the radial coordinate $r$. 
\eqs{e13} and \eqref{e14} coincide exactly with the corresponding relativistic equations, whereas the only difference with respect to GR in \eq{e12} is a factor $\cc^2/\mathcal{E}\simeq1$ multiplying the first term on its right hand side. 

For the actual implementation of the present approach within a numerical code, it can sometimes be convenient to rewrite \eqs{e12}-\eqref{e14} in terms of Cartesian coordinates. The resulting acceleration is then
\begin{equation}
\ddot{x}^{i} = -x^i\frac{\alpha^2\Phi'}{r}+\dot{x}^{i}\dot{r}\,\frac{\alpha'}{\alpha}-x^i\dot{\varphi}^2\left(1-\alpha+\frac{r\,\alpha'}{2}\right),
\label{e15}
\end{equation}
with ${x}^{i}=\{x,y,z\}$. In \eq{e15}, $\dot{\varphi}^2$ and $\dot{r}$ are calculated in terms of Cartesian coordinates as
\begin{gather}
\dot{\varphi}^2 = \frac{(x\,\dot{y}-y\,\dot{x})^2+(z\,\dot{y}-y\,\dot{z})^2+(x\,\dot{z}-z\,\dot{x})^2}{r^4},
\label{e16}\\
\dot{r} = \frac{x\,\dot{x}+y\,\dot{y}+z\,\dot{z}}{r}.
\label{e17}
\end{gather}

In Paper~I we demonstrated the simplicity of implementing \eq{e15} within a Newtonian smoothed particle hydrodynamics code. We applied this code for studying the tidal disruption of a solar-type star by a supermassive black hole and found a very good agreement with previous relativistic simulations. Even though this was done for the particular case of Schwarzschild spacetime, there is nothing specific to this metric that may impede an analogous implementation of the present approach for any other static, spherically symmetric black hole spacetime.

\section{Circular orbits}
\label{S4}

Circular orbits are defined by the conditions $\dot{r} = 0$, $\ddot{r}= 0$. Using \eqs{e10} and \eqref{e12}, and solving the corresponding system of two equations for $E_*$ and $h_*$, we get
\begin{gather}
h_*^c = \sqrt{\frac{2\,r^3\Phi' }{2\,\alpha-r\,\alpha'}},
\label{e18}\\
E_*^c = \Phi+\frac{r\,\alpha\,\Phi' }{2\,\alpha-r\,\alpha'},
\label{e19}
\end{gather}
where the superscript $c$ is used to indicate that these expressions apply to circular orbits only. 

It is simple to verify that both $h_*^c$ and $E_*^c$ coincide with their exact relativistic counterparts, i.e.~$h^c=h_*^c$ and $E^c=E_*^c$. This has the important consequence that the radial locations of all of the distinct circular orbits are captured exactly: the photon circular orbit is found at the location where both \eqs{e18} and \eqref{e19} diverge to infinity (i.e.~where $r\,\alpha'=2\,\alpha$), the marginally bound circular orbit is defined by the condition $E_*^c = 0$, and the marginally stable circular orbit (also called the innermost stable circular orbit -- ISCO) is located at the point at which both $h_*^c$ and $E_*^c$ reach their minima. This last condition is equivalent to finding the smallest positive root of
\begin{equation}
\mathcal{A}(r) =
3\,\alpha-2\,r\,\alpha'+r\,\alpha\,\frac{\alpha''}{\alpha'} = 0.
\label{e20}
\end{equation}
For some particular spacetimes, the condition $\mathcal{A}(r)=0$ might have more than one positive root or no real roots at all. See, e.g.~\cite{stuchlik09} for a discussion of this in the case of Schwarzschild--de Sitter spacetime.

\begin{figure}
\begin{center}
  \includegraphics[width=120mm]{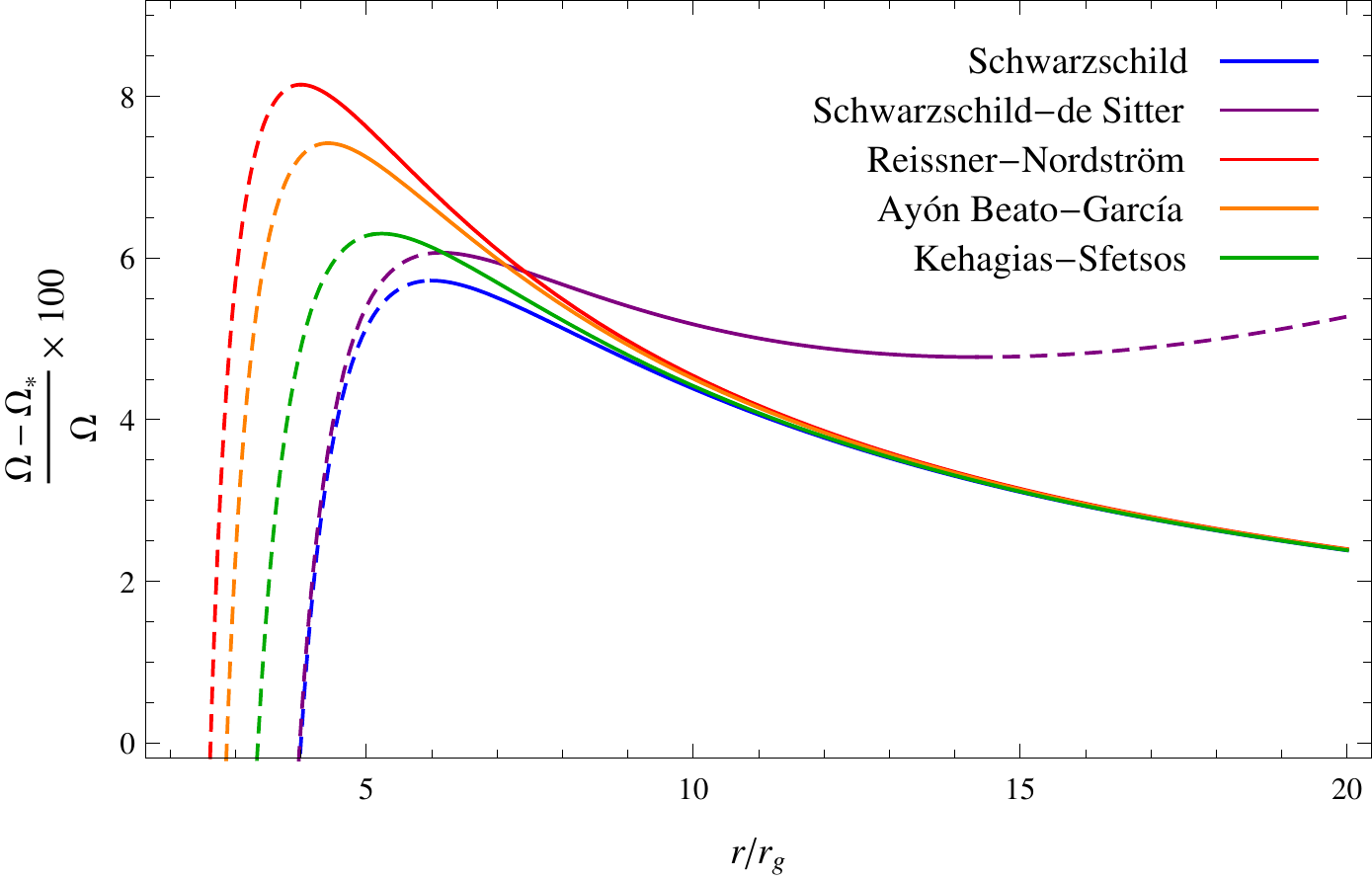}
\end{center}
\caption{Relative error with which the Keplerian and epicyclic frequencies are reproduced by our generalized Newtonian description for the examples of static, spherically symmetric spacetimes listed in Table~\ref{t1}. Note that the relative error on the vertical axis satisfies \mbox{$(\Omega-\Omega_*)/\Omega = (\Omega^\parallel-\Omega_*^\parallel)/\Omega^\parallel =1-\xi$} given that both $\Omega_*$ and $\Omega_*^\parallel$ conform to \mbox{$\Omega_*= \xi\,\Omega$} and \mbox{$\Omega_*^\parallel = \xi\,\Omega^\parallel$}. A continuous trace on each curve represents stable circular motion while a broken line corresponds to unstable circular motion. Note that the boundaries of stability correspond to extreme points of $\xi$. For Schwarzschild-de Sitter spacetime we have taken $\Lambda\,\rg^2 = 2\times10^{-4}$, for Reissner--Nordstr\"om spacetime $q=\rg$, for Ay\'on-Beato--Garc\'ia spacetime $q=q_\s{\text{crit}} \simeq 0.634\,\rg$, and for Kehagias-Sfetsos $\omega\,\rg^2 = 1/2$ (see Table~\ref{t1} for further details about the meaning of these parameters).}
\label{f1} 
\end{figure}

The orbital or Keplerian angular frequency of a circular orbit $(\Omega_*\equiv \dot{\varphi}^c)$ can be calculated from \eqs{e9} and \eqref{e18} as
\begin{equation}
\Omega_*= 
\sqrt{\frac{2\,\alpha^2\Phi' }{r(2\,\alpha-r\,\alpha')}} = \xi\, \Omega, 
\label{e21}
\end{equation}
where 
\begin{equation}
\xi = \sqrt{\frac{2\,\alpha^2}{2\,\alpha-r\,\alpha'}}
\label{e22}
\end{equation}
gives the deviation of our approach from the exact relativistic value \mbox{$\Omega = \sqrt{\Phi'/r}$}. Note that the condition in \eq{e20} is equivalent to finding the extrema of $\xi$, i.e.~$\mathcal{A}(r)=0$ $\iff$ $\ud\xi/\ud r = 0$. In Figure~\ref{f1} we have plotted the relative error $(\Omega-\Omega_*)/\Omega=1-\xi$ for the examples of spherically symmetric black hole solutions listed in Table~\ref{t1}. For the cases considered in this figure, we see that the percentage error associated with the present generalized Newtonian description of the Keplerian frequency is less than $10\%$ for all radii for which stable circular motion is possible. 

\section{Epicyclic frequencies}
\label{S5}

In this section we consider perturbations away from a stable circular orbit. Without loss of generality, we can assume that the unperturbed trajectory lies on the equatorial plane. Let us denote by $\delta r$, $\delta\theta$, and $\delta\phi$ small perturbations in the radial, polar, and azimuthal directions, respectively. To first order, the equations governing the perturbed trajectory are obtained from \eqs{e12}-\eqref{e14} as
\begin{align}
\delta\ddot{r}  = & -r\left(2\,\alpha-r\,\alpha'\right)\Omega_*
\left(\Omega_*'\,\delta r -\delta\dot{\phi}\right) , 
\label{e23} \\
\delta\ddot{\theta} = & -\Omega_*^2\delta\theta,
\label{e24}  \\
\delta\ddot{\phi} = & -\frac{\left(2\,\alpha-r\,\alpha'\right)\Omega_*}{r\,\alpha}\,\delta \dot{r},
\label{e25} 
\end{align}
where $\Omega_*$ is the Keplerian angular frequency of a circular orbit given in \eq{e21}.

From \eq{e24} we immediately see that perturbations orthogonal to the orbital plane (i.e.~along the polar direction) are decoupled from those parallel to it and, furthermore, that the  associated epicyclic frequency satisfies
\begin{equation}
\Omega_*^\perp = \Omega_*.
\label{e26} 
\end{equation}
On the other hand, \eqs{e23} and \eqref{e25} show that perturbations parallel to the orbital plane (i.e.~along the radial and azimuthal directions) are coupled together. If we assume that these two oscillatory modes share a common frequency $\Omega_*^\parallel$ and follow the same procedure as in Paper~I, we get
\begin{equation}
\Omega_*^\parallel = \sqrt{\mathcal{A}(r)}\,\Omega_*,
\label{e27} 
\end{equation}
with $\mathcal{A}$ as defined in \eq{e20}.

When we compare \eqs{e26} and \eqref{e27} with the corresponding relativistic expressions, we find that the departures from GR are given by the same function $\xi$ defined in \eq{e22} for the Keplerian frequency $\Omega_*$, i.e.
\begin{equation}
\Omega_*^\perp = \xi\,\Omega^\perp 
\quad\text{and}\quad
\Omega_*^\parallel = \xi\,\Omega^\parallel.
\end{equation}
From here we arrive at the important result that the two epicyclic frequencies $\Omega_*^\perp$ and $\Omega_*^\parallel$ keep the same ratio with respect to $\Omega_*$ as the corresponding relativistic ones, i.e.
\begin{equation}
\frac{\Omega_*^\perp}{\Omega_*} = \frac{\Omega^\perp}{\Omega} = 1
\quad\text{and}\quad
\frac{\Omega_*^\parallel}{\Omega_*} = \frac{\Omega^\parallel}{\Omega} = 
\sqrt{\mathcal{A}(r)}.
\end{equation}

In Figure~\ref{f1} we plot the relative error with which the Keplerian and epicyclic frequencies are reproduced with the new prescription for the examples of spherically symmetric spacetimes in Table~\ref{t1}.

\newpage

\section{Summary}
\label{S6}

We have presented a generalized Newtonian description of the motion of test particles valid for any static, spherically symmetric spacetime based on the low-energy limit of geodesic motion. This work constitutes a generalization of the approach developed in Paper~I for a Schwarzschild black hole. Among the various examples of spherically symmetric spacetimes that might be of relevance for different astrophysical contexts, we have considered the use of the new approach for a Schwarzschild black hole with cosmological constant (Schwarzschild--de Sitter spacetime), a charged black hole in Einstein--Maxwell theory (Reissner--Nordstr\"om spacetime), a charged regular black hole in Einstein theory coupled with a non-linear electrodynamics (Ay\'on-Beato--Garc\'ia spacetime), and a black hole solution of the Lorentz-violating gravitational theory of Ho{\v r}ava-Lifshitz (Kehagias-Sfetsos spacetime).

We have shown that the new approach reproduces exactly the radial location of the photon, marginally bound, and marginally stable circular orbits. Moreover, the radial dependence of the binding energy and angular momentum of circular orbits is also reproduced exactly. In addition, the spatial projection and the pericentre shift of general particle trajectories coincide exactly with the corresponding relativistic expressions. On the other hand, we have shown that, for the spacetimes considered in this paper, the Keplerian and epicyclic frequencies are reproduced to better than $10\%$ for all of the radii that allow stable circular motion. 

We believe that the ability of the present approach to reproduce all of these key relativistic features simultaneously and, in addition, its ease of implementation within existing Newtonian frameworks, make it a simple, yet powerful tool for studying astrophysical accretion flows onto a broad family of black hole spacetimes.

\ack

We would like to thank John C.~Miller and Zdenek Stuchl\'ik for insightful discussions and critical comments on the manuscript. Part of this work was started during the conference Prague Synergy 2013. ET would like to thank the organizers for their hospitality and the participants for stimulating discussions. This work has been supported by the Swedish Research Council (VR) under the grant 621-2012-4870.

\begin{landscape}
\begin{table}
\caption{Application to examples of static and spherically symmetric black hole spacetimes. The function $\alpha$ is connected with the spacetime metric as in \eq{e1}. The ratio $\Omega_*/\Omega = \Omega_*^\parallel/\Omega^\parallel$ corresponds to the function $\xi$ defined in \eq{e22}. The second to last column gives the maximum percentage error obtained for $\Omega_*$, $\Omega_*^\parallel$ and $\Omega_*^\perp$ (we have considered only values of $r$ for which circular motion is stable, see Figure~\ref{f1}).} 
\label{t1}
\begin{tabular}{llllp{5.7cm}}
\br\\[-7pt]
Spacetime & $\alpha$ & $\frac{\Omega_*}{\Omega}$ & $\frac{\Omega-\Omega_*}{\Omega}$  & Notes
\\[6pt]\mr\\[-7pt]
Schwarzschild & $1 - \frac{ 2\,\rg }{ r }$ & 
$\frac{r-2\,\rg}{\sqrt{r(r-3\,\rg)}}$ & $\leqslant 5.7\%$ & $\rg = \G M/\cc^2$ is the gravitational radius and $M$ is the mass of the black hole.\\ 
Schwarzschild--de Sitter
& $1 - \frac{ 2\,\rg }{ r } - \frac{\Lambda\,r^2}{3}$ & 
$\frac{r-2\,\rg-\Lambda\,r^3/3}{\sqrt{r(r-3\,\rg)}}$ & $\leqslant 7.1\%$ & $\Lambda$ is the cosmological constant. The condition $\Lambda\,\rg^2 \leqslant 7.11\times10^{-4}$ should be satisfied in order to have stable circular orbits. \\ 
Reissner--Nordstr\"om & $ 1 - \frac{ 2\,\rg }{ r } +\frac{\rQ^2}{r^2}$ & 
$\frac{r^2-2\,\rg\,r+\rQ^2}{r\sqrt{r^2-3\,\rg\,r+2\,\rQ^2}}$ & $\leqslant 8.2\%$ & $\rQ = \sqrt{\G}\,Q/\cc^2$ and $Q$ is the electrical charge of the black hole. To prevent the formation of a naked singularity, the condition $|\rQ| < \rg$ should be satisfied.\\ 
Ay\'on-Beato--Garc\'ia & $ 1 -\frac{ 2\,\rg \,r^2}{\left(r^2+\rQ^2\right)^{3/2} }+\frac{ \rQ^2r^2}{\left(r^2+\rQ^2\right)^{2} }$ & 
$\frac{1 -\frac{ 2\,\rg \,r^2}{\left(r^2+\rQ^2\right)^{3/2} }+\frac{ \rQ^2r^2}{\left(r^2+\rQ^2\right)^{2} }}{\left[1 -\frac{ 3\,\rg \,r^4}{\left(r^2+\rQ^2\right)^{5/2} }+\frac{ 2\,\rQ^2r^4}{\left(r^2+\rQ^2\right)^{3} }\right]^{\nicefrac{1}{2}}}$ & $\leqslant 7.5\%$ & We have considered values of $q\leqslant q_\s{\text{crit}} \simeq 0.634\,\rg$ in order to ensure a monotonic dependence of the ISCO on $q$. \\ 
Kehagias-Sfetsos & $ 1 +\omega\,r^2\left[
1-\left(1+\frac{4\,\rg}{\omega\,r^3}\right)^{\nicefrac{1}{2}}\right]$ & 
$\frac{1 +\omega\,r^2\left[
1-\left(1+\frac{4\,\rg}{\omega\,r^3}\right)^{\nicefrac{1}{2}}\right]}{\left[1-\frac{3\,\rg}{r}\left(1+\frac{4\,\rg}{\omega\,r^3}\right)^{-\nicefrac{1}{2}}\right]^{\nicefrac{1}{2}}}$ & $\leqslant 6.3\%$ & $\omega$ gives the deviation away from GR (which is recovered in the limit \mbox{$\omega\rightarrow \infty$}). The condition $\omega\,\rg^2\geqslant 1/2$ should be satisfied to prevent the emergence of a naked singularity.\\
\br
\end{tabular}
\end{table}
\end{landscape}

\bibliographystyle{jphysicsB}
\bibliography{references}

\end{document}